\documentclass[%
reprint,
 amsmath,amssymb,
 aps,
 prx,
twocolumn,
superscriptaddress,
longbibliography
]{revtex4-2}

\usepackage{graphicx}
\usepackage{dcolumn}
\usepackage{bm}
\usepackage{mathrsfs} 
\usepackage{lmodern}

 \ifx\pdftexversion\undefined
\usepackage[dvips]{hyperref}
\else
\usepackage{hyperref}
\fi
\hypersetup{
  colorlinks = true, linkcolor = blue
}

\begin{document}

\preprint{APS/123-QED}

\title{Tomography of a single-atom-resolved detector in the presence of shot-to-shot number fluctuations}

\author{Maxime Allemand}
\affiliation{Université Paris-Saclay, Institut d’Optique Graduate School,
CNRS, Laboratoire Charles Fabry, 91127, Palaiseau, France}
\author{Raphael Jannin}
\affiliation{Université Paris-Saclay, Institut d’Optique Graduate School,
CNRS, Laboratoire Charles Fabry, 91127, Palaiseau, France}
\author{Géraud Dupuy}
\affiliation{Université Paris-Saclay, Institut d’Optique Graduate School,
CNRS, Laboratoire Charles Fabry, 91127, Palaiseau, France}
\author{Jan-Philipp Bureik}
\affiliation{Université Paris-Saclay, Institut d’Optique Graduate School,
CNRS, Laboratoire Charles Fabry, 91127, Palaiseau, France}
\author{Luca Pezzè}
\affiliation{Istituto Nazionale di Ottica del Consiglio Nazionale delle Ricerche (CNR-INO), Largo Enrico Fermi 6, 50125 Firenze, Italy}
\affiliation{European Laboratory for Nonlinear Spectroscopy (LENS), Via N. Carrara 1, 50019 Sesto Fiorentino, Italy}
\affiliation{QSTAR, Largo Enrico Fermi 2, 50125 Firenze, Italy}
\author{Denis Boiron}
\affiliation{Université Paris-Saclay, Institut d’Optique Graduate School,
CNRS, Laboratoire Charles Fabry, 91127, Palaiseau, France}
\author{David Clément}
\affiliation{Université Paris-Saclay, Institut d’Optique Graduate School,
CNRS, Laboratoire Charles Fabry, 91127, Palaiseau, France}

\date{\today}

\begin{abstract}
Tomography of single-particle-resolved detectors is of primary importance for characterizing particle correlations with applications in quantum metrology, quantum simulation and quantum computing. However, it is a non-trivial task in practice due to the unavoidable presence of noise that affects the measurement but does not originate from the detector. 
In this work, we address this problem for a three-dimensional single-atom-resolved detector where shot-to-shot atom number fluctuations are a central issue to perform a quantum detector tomography. We overcome this difficulty by exploiting the parallel measurement of counting statistics in sub-volumes of the detector, from which we evaluate the effect of shot-to-shot fluctuations and perform a local tomography of the detector. In addition, we illustrate the validity of our method from applying it to Gaussian quantum states with different number statistics. Finally, we show that the response of Micro-Channel Plate detectors is well-described from using a binomial distribution with the detection efficiency as a single parameter.
\end{abstract}

\maketitle

\section{Introduction}

Many-body correlation functions and full counting statistics (FCS) unveil information that is not contained in one-body quantities and statistical averages. This information is crucial for understanding strongly-correlated systems, a major goal for quantum simulators \cite{Bloch2008, Amico2012, Schweigler2017, Georgescu2014, Daley2022}. 
Moreover, many-body correlations may be used as ressources to enhance performances in quantum sensing \cite{Giovannetti2011, Ji2024} and in quantum computing \cite{Demirel2021}.
FCS are also relevant in quantum sensing \cite{Crooker2004, Degen2017, Pezze2018} by providing direct access to the Fisher information and thus to useful entanglement properties of a quantum state \cite{Strobel2014, Pezze2009, Hyllus2012, Geza2012}.
One important challenge to access FCS is the accurate characterization of the measurement observables, particularly due to the unavoidable detection noise.
The full characterization of a detector \textit{i.e.} determining the detector response to any quantum state is generally referred to as quantum detection tomography (QDT) \cite{Luis1999, Fiurasek2001, Feito2009, Laurat2011}. Such characterization does not rely on any assumption, contrary to describing a detector with a single parameter like its detection efficiency.

QDT methods consist in determining a set of positive-operator-valued measures (POVM) that fully characterizes the measurement results \cite{Nielsen2010, Gebhart2023}.
In the case of single-particle detectors, the POVMs \{$\Pi_n$\} relate the measured probabilities $P(N^\mathrm{det}=n)$ to the density matrix $\rho$:
\begin{equation}
\label{eq: POVMs}
    P(N^\mathrm{det}=n) = \mathrm{Tr} \left( \Pi_n \rho \right), \ \ \ \forall n \in \mathbb{N}
\end{equation}
where $P(N^\mathrm{det}=n)$ is the probability that the detected number $N^\mathrm{det}$ is equal to an integer value $n$. A perfect single-particle detector is characterized by $\Pi_n=|n\rangle \langle n|$ and it measures the diagonal elements of $\rho$ in the basis of Fock states.

Equation~\eqref{eq: POVMs} has a significant implication for tomography methods: to evaluate the POVMs \{$\Pi_n$\} from the measured counting statistics $P(N^\mathrm{det}=n)$, one needs to know precisely the full density matrix $\rho$, a task challenging to achieve in practice. 
In addition, fluctuations in $\rho$ arising from the production of the probed state -- and thus not associated to the detection process -- could be detrimental to QDT approaches. In photonics, this issue is often minor since intensity fluctuations of continuous-wave lasers are small and minimally impact  the photon number statistics. As a matter of fact, QDT has mainly been used for the counting of photons \cite{Lundeen2009, Zhang2012, Brida2012, Zhang2020} and the characterisation of homodyne detectors \cite{Grandi2017}. Only recently  has it been applied to qubit readout for pairs of trapped ions \cite{Keith2018}, non-demolition measurement of superconducting qubits \cite{Chen2019}, and high-fidelity reconstruction of states \cite{Hetzel2022}. In contrast to photonics, cold samples of atoms or molecules are produced in a pulsed way and with large shot-to-shot number fluctuations (typically a few percent of its mean value). Applying QDT methods in these platforms necessitates accounting for these number fluctuations.

In this work, we introduce a method for characterizing the response of single-particle detectors in the presence of shot-to-shot fluctuations. Our method applies to detectors comprising many elementary detection units operating in parallel, such as scientific cameras with their pixels or Micro-Channel Plates (MCP) with their channels. We exploit the possibility to perform two measurements: over the entire detector -- to monitor shot-to-shot fluctuations -- and in sub-volumes (pixels or voxels) -- to perform a local tomography of the detector. Furthermore, we show 
that adequate choices on quantum states and sub-volumes enable us to deduce information about the density matrix $\rho$ effectively probed by the detector. This resolves the challenge of distinguishing the two terms $\Pi_n$ and $\rho$ when only their product is measured, see Eq.~\eqref{eq: POVMs}, in contrast to previous works \cite{Keith2018, Hetzel2022}. We demonstrate the validity of our method by characterizing a detector of metastable helium atoms (He$^*$) using two states whose statistics strongly differ, a Bose-Einstein condensate (BEC) and a Mott insulator \cite{Carcy2019, Cayla2020, Herce2023}. Finally, our findings confirm that MCPs accurately measure the full counting statistics, with applications to the detection of various atomic samples ranging from noble gases \cite{Vassen2012}, ions \cite{Veit2021} and Rydberg atoms \cite{Zuber2022} to molecules \cite{Yu2022}.

\section{Tomography in the presence of shot-to-shot number fluctuations}

Single-particle detectors are phase-insensitive. This implies that the POVMs \{$\Pi_i$\} are diagonal in the Fock states basis. They write $\Pi_i = \sum_{j=0}^{l-1} V_{ij} | j \rangle \langle j |$ where $V_{ij}=P(N^\mathrm{det}=i|N=j)$ is the conditional probability of detecting $i$ particles given that $j$ particles arrive on the detector. Here the Hilbert space of Fock states is truncated to a dimension $l$ corresponding to the maximum number of particles in the system. 
Using Eq.~\eqref{eq: POVMs}, the coefficients $V_{ij}$ can be determined from measuring the counting statistics of $l$ different states and from inverting the $l$ associated equations: 
\begin{equation}
\label{eq: V}
    P(N^\mathrm{det}=i) = \sum_{j=0}^{l-1} V_{ij} P(N=j),
\end{equation}
for $i = 1,\ ...,\ l $, where $P(N=j) = \langle j | \rho | j \rangle$ are the statistics of the state, \textit{i.e.} before detection. A perfect detector is characterized by $V_{ij} = \delta_{ij}$, i.e. a  matrix $V$ of the coefficients $V_{ij}$ which is the identity matrix.

To obtain the matrix $V$ for a non-perfect detector, one possibility is to use Fock states and construct the POVMs of the detector from Eq.~(\ref{eq: V}) at each value of $0 \leq j \leq l-1$. However, producing Fock states is extremely challenging, if not impossible, for most experimental platforms. In addition, establishing the production of Fock states requires a precise knowledge of the response of the detector. 
An alternative choice consists in probing states whose statistics are simple to characterize using an independent device. For instance, the coherent state produced by a laser well-above threshold is fully characterized by its average intensity $\langle I \rangle$ which is easily measured with a photodetector. The tomography ({\it i.e.} determining the POVMs $\Pi_i$) can thus be decoupled from the statistics of the light ({\it i.e.} knowing the probabilities $P(N=n)$ from the measured value of $\langle I \rangle$). This protocol was used for performing the QDT of single-photon counters in Ref.~\cite{Lundeen2009}.
Beyond coherent states, the use of Gaussian states whose statistics depend on a single parameter $\mu$, the mean particle number $\mu=\langle N \rangle$, are well suited to QDT. With such Gaussian states, shot-to-shot fluctuations can be described from considering $\mu$ as a fluctuating variable. The statistics of the state probed by the detector are indeed the composition of the bare state statistics with the statistics of the fluctuating parameter $\mu$. 

\begin{figure}
\includegraphics[width=\columnwidth]{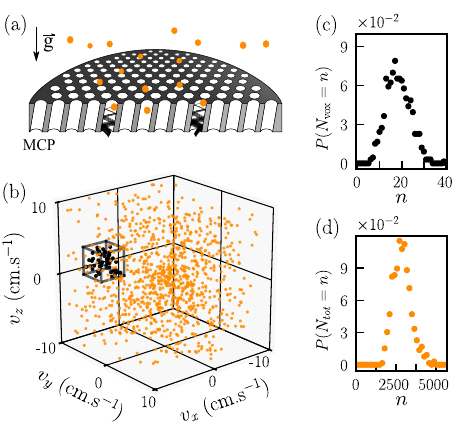}
\caption{\label{fig:detector} (a) Illustration of the detection with a Micro-Channel Plate (MCP). $^4$He$^*$ atoms impinge the MCP detector and trigger electron avalanches resulting in detection events. (b) Reconstructed 3D atom distribution in a single experimental shot. Each dot is a detection event of a $^4$He$^*$ atom with velocity $(v_x, v_y, v_z)$. The black box depicts one of the cubic voxels where the tomography is performed (see text). (c) Counting statistics of the atom number $N_\mathrm{vox}$ in a voxel identified as a black box in panel (b). The binsize is 1. (d) Counting statistics of the total atom number $N_\mathrm{tot}$. The binsize is 155.}
\end{figure}

A central result of our work is to identify an approach to measure the counting statistics of the mean particle number $\langle N \rangle $ and to predict the statistics $P(N=n)$ to be used in Eq.~\eqref{eq: V}. 
We exploit the fact that, in most situations of interest, the correlation length $l_c$ of the bare state -- i.e. the distance beyond which two particles are essentially uncorrelated -- is much smaller than the size $D$ of the detector. A sub-volume of the detector of intermediate size $l_c \ll d \ll D$, named voxel in the following, probes many correlation volumes of the quantum state. This choice implies that most of the atoms in the voxel are uncorrelated with each other, with the consequence that the number statistics in the voxel are Poisson statistics, regardless of the statistical properties of the Gaussian state we consider.
Our method applies to detectors made of a large number of elementary detectors working in parallel: the shot-to-shot fluctuations are monitored from using the entire detector and this information is used to determine the expected statistics $P(N_\mathrm{vox}=n)$ of the particle number $N_\mathrm{vox}$ in voxels. The tomography of the detector is performed from analysing its response in these voxels.\\

Our experiment produces ultracold gases of metastable Helium-4 ($^4\mathrm{He}^*$), with unavoidable shot-to-shot fluctuations of the total atom number \cite{Bouton2015}. These atomic clouds are detected destructively by letting them fall on a two-dimensional MCP detector illustrated on Fig.~\ref{fig:detector}(a). The MCP is composed of about $2 \times 10^7$ elementary detectors (micro-channels) working in parallel \cite{Hoendervanger_2013, Jagutzki2002}. 
The positions and arrival times of the atoms on the detector enable the atom-by-atom reconstruction of the 3D momentum distribution of the gas as shown in Fig.~\ref{fig:detector}(b). 
An example of the counting statistics obtained in a voxel (resp. over the full detector) is shown in Fig.~\ref{fig:detector}(c) (resp. Fig.~\ref{fig:detector}(d)). We use a voxel whose size comprises several elementary detectors (several micro-channels in the detector plane and several time bins out-of-plane).
Below we apply our tomography method to two types of quantum states produced in our experiment, a Bose-Einstein condensate (BEC) and a Mott insulator, whose statistics are markedly different. The statistics of the BEC is that of a coherent state (for the data presented here we measure $g^{(2)}(0) = 1.07(6)$ using the method described in \cite{Herce2023}), while the statistics of the Mott state are those of a Gaussian thermal state in momentum space (see Fig.~\ref{fig:thermal distribution} in Appendix~\ref{app:A}). In the following, we first describe our tomography method in the case of Mott insulators. Then we show its results and compare them with those obtained in the case of BECs.

\section{Model of the expected counting statistics prior to detection}

We calculate the expected counting statistics of the atom number falling into a voxel by accounting {\it (i)} for the statistics of the quantum state (prior to detection) and {\it (ii)} for the shot-to-shot fluctuations. As discussed above, we use quantum states whose statistics are determined by a single parameter $\mu=\langle N \rangle$. 
In the Mott case, we choose on purpose the volume of a voxel such that it contains many ($\sim125$) correlation volumes, where a correlation volume is defined as the volume over which bosons exhibit bunching after time-of-flight \cite{Herce2023}. As discussed above, this choice implies that the atom number in a voxel $N_\mathrm{vox}$ follows Poisson statistics: 
\begin{equation}
\label{eq:Poisson}
P(N_\mathrm{vox}=n | N_\mathrm{tot}=\lambda) = e^{-\mu} \frac{\mu^n}{n!},
\end{equation}
where $\mu=\langle N_\mathrm{vox} \rangle |_{N_\mathrm{tot}=\lambda}$ is the conditional mean of $N_\mathrm{vox}$ for a given value of the total atom number $N_\mathrm{tot}=\lambda$.

We account for shot-to-shot atom number fluctuations by assuming that they result in fluctuations of the atomic (momentum) density by a global factor. The mean atom number in the voxel $\langle N_\mathrm{vox} \rangle |_{N_\mathrm{tot}=\lambda}$ is proportional to the total atom number $N_\mathrm{tot}=\lambda$, i.e.  $\langle N_\mathrm{vox} \rangle|_{N_\mathrm{tot}=\lambda} = \alpha \times \lambda$ where $\alpha = \langle N_\mathrm{vox} \rangle / \langle N_\mathrm{tot} \rangle$. 
When the total atom number $N_\mathrm{tot}$ fluctuates, $N_\mathrm{vox}$ follows Poisson statistics with a fluctuating parameter:
\begin{equation}
\label{eq:distribution}
\begin{split}
P(N_\mathrm{vox}=n) &= \sum_{\lambda \in \mathbb{N}} P(N_\mathrm{tot}=\lambda) P(N_\mathrm{vox}=n | N_\mathrm{tot}=\lambda)\\
    &
    = \sum_{\lambda \in \mathbb{N}} P(N_\mathrm{tot}=\lambda) e^{-\alpha\lambda} \frac{(\alpha\lambda)^n}{n!}.
\end{split}
\end{equation} 

\begin{figure}[t!] 
\includegraphics{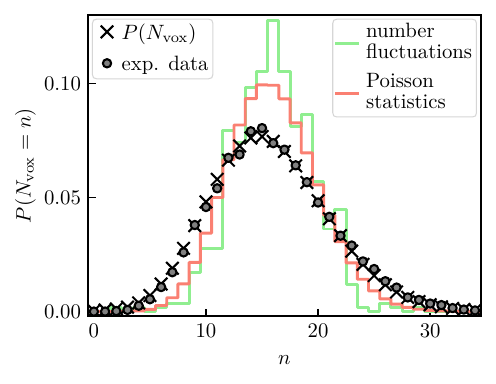} 
\caption{\label{fig:distribution} Counting statistics $P(N_\mathrm{vox}=n)$ of the atom number per voxel $N_\mathrm{vox}$ recorded with Mott insulators. The dots are experimental data and the crosses are the prediction of Eq.~\eqref{eq:distribution}. 
For illustration, the two contributions to the predicted statistics in Eq.~\eqref{eq:distribution} are shown as solid lines: Poisson statistics with parameter $\mu = \langle N_\mathrm{vox} \rangle$ associated with the probed state (red) and fluctuations of the total atom number rescaled by $\alpha = \langle N_\mathrm{vox} \rangle / \langle N_\mathrm{tot} \rangle$ (green).}
\end{figure}

To estimate in the experiment the two contributions to the expected statistics (see Eq.~\eqref{eq:distribution}), we analyse a set of $\sim 600$ shots recorded in identical conditions. The counting statistics of the total atom number $P(N_\mathrm{tot}=\lambda)$ are measured directly by counting atoms impinging the full detector. To increase the signal-to-noise ratio of the counting statistics in a voxel, we make use of the spherical symmetry of the atomic clouds. The counting statistics $P(N_\mathrm{vox}=n)$ shown in Fig.~\ref{fig:distribution} are an average of the counting statistics of 24 voxels with identical atomic densities (see Appendix~\ref{App:B}). 
The prediction from Eq.~\eqref{eq:distribution} is in excellent agreement with the measured counting statistics. Fig.~\ref{fig:distribution} also illustrates the need to account for both contributions -- from shot-to-shot fluctuations and from quantum statistics -- as they exhibit similar widths.

\section{Detector tomography using Mott insulators}

\subsection{Counting statistics with a varying mean atom number}

We proceed to perform the tomography of the detector, i.e. determine the coefficients $V_{ij}$ shown in Eq.~\eqref{eq: V}, using the knowledge of the expected counting statistics $P(N_\mathrm{vox}=n)$ in the voxel (see Eq.~\eqref{eq:distribution}). Experimentally, one has to sample the mean detected atom number to span a large range of values and access a large set of POVMs.
To this aim, we produce the same Mott insulator state and sample different atom numbers by sending a controlled fraction of the atoms towards the detector 
\footnote{In practice, we prepare the atomic clouds in the spin-polarized $|2^3\mathrm{S}_1, m_J = +1\rangle$ state before transferring a controlled fraction to the $|2^3\mathrm{S}_1, m_J = 0\rangle$ state towards the detector (the remaining atoms in the magnetic sublevel $|m_J = +1\rangle$ are expelled from the detector area using a magnetic gradient). This transfer is achieved using a two-photon Raman transition with co-propagating beams \cite{Tenart2021} and the number of atoms sent to the detector is varied by controlling the duration $t$ of the associated Rabi oscillations.}, as introduced in Ref.~\cite{Hetzel2022}. Varying the duration $t$ of the associated Rabi oscillations samples the mean detected atom number.
Importantly, the Poissonian nature of the counting statistics measured in a voxel does not vary with $t$. Indeed, the Rabi oscillation acts as a beam splitter which selects atoms according to a binomial distribution of efficiency $\sin^2(\pi f_R t)$ independently on the atom momentum. For each duration $t$, the resulting counting statistics $P(N_\mathrm{vox}=n)$ are Poissonian with the mean rescaled by $\sin^2(\pi f_R t)$.

\begin{figure}[h!]
\includegraphics{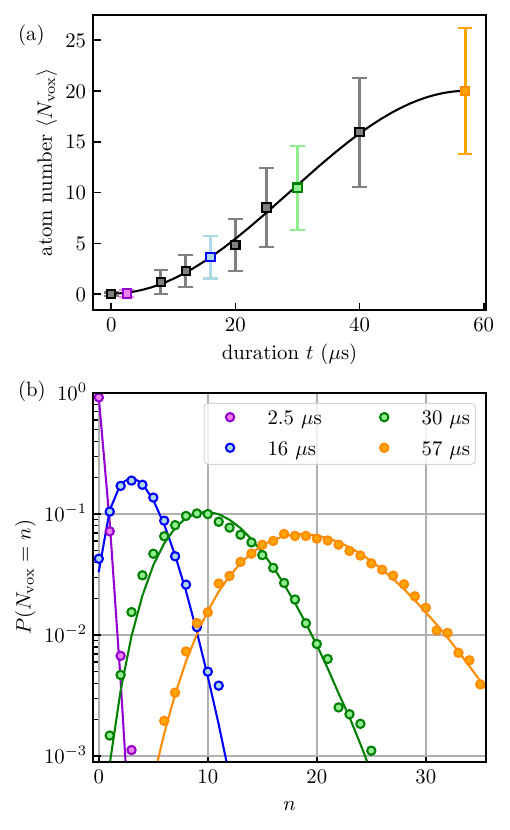}
\caption{\label{fig:data} (a) Measured mean atom number $\langle N_{\rm vox} \rangle$ plotted as a function of the duration of the Rabi pulse (experimental data using Mott insulators). The markers indicate the mean value and the error bars indicate one standard deviation. The black line is a sinusoidal fit of the mean values. (b) Counting statistics of $N_{\rm vox}$ at different pulse duration $t$ for the data shown in (a). Circles are the experimental data and solid lines are the predictions from Eq.~\eqref{eq:algorithm params}.
}
\end{figure}

We record counting statistics at varying Rabi pulse durations \{$t_k$\}$_{k=1..10}$, with their mean atom numbers $\langle N_{\rm vox} \rangle_k$ plotted in Fig.~\ref{fig:data}(a) as a function of the Rabi duration. We fit the variation of $\langle N_{\rm vox} \rangle$ with  $\langle N_\mathrm{vox} \rangle_k = n_\pi\sin^2(\pi f_R t_k)$ to extract the mean atom number $n_\pi$ at $\pi$-pulse and the Rabi frequency $f_R$.

From using Eq.~\eqref{eq:distribution}, the expected counting statistics at any duration $t_{k}$ writes: 
\begin{equation}
\label{eq:algorithm params}
P_{k}(N_\mathrm{vox}=n) = \sum_{\lambda \in \mathbb{N}} P(N_\mathrm{tot}=\lambda) e^{-\alpha(t_{k})\lambda} \frac{(\alpha(t_{k})\lambda)^n}{n!}
\end{equation}
where $\alpha(t_{k}) = n_\pi\sin^2(\pi f_R t_{k}) / \langle N_\mathrm{tot} \rangle$.
Examples of measured counting statistics together with the corresponding predictions $P_{k}(N_\mathrm{vox}=n)$ obtained from Eq.~\eqref{eq:algorithm params} are shown in Fig.~\ref{fig:data}(b). 
Only a few input parameters are used to compute the expected counting statistics $P_{k}$: $n_\pi$ and $f_R$ which are fitted from Fig.~\ref{fig:data}(a) and the counting statistics of the total atom number which are measured -- in parallel of the voxel atom number -- at the $\pi$-pulse. The overall agreement over the full range of atom numbers sampled in the experiment validites our approach to predict the counting statistics in a voxel using Eq.~\eqref{eq:algorithm params}. Note that this comparison implicitly assumes a perfect detector.  Below, we include the effect of an imperfect detector through the matrix $V$ introduced in Eq.~\eqref{eq: V}.

To further increase the number of independent samples, we exploit the density profile of the atomic cloud and use two additional voxels located at different atomic densities (also averaged using spherical symmetry). This effectively amounts to sample in parallel 3 Rabi oscillations whose average atom numbers differ, leading to $3 \times 10$ measured counting statistics.

\subsection{Optimization procedure}

The tomography consists in comparing the expected counting statistics from Eq.~\eqref{eq:algorithm params} with the detected ones and in attributing the discrepancies to the imperfections of the detector. We emphasize that the tomography considers all the experimental imperfections -- except the atom number fluctuations -- as detector imperfections. Such imperfections include e.g. fluctuations of the Rabi frequency and statistical noise. Therefore, the tomography outcome depicts the detector's performance in the worst-case scenario.

We estimate $V$ from minimizing the sum of the differences between measured and reconstructed statistics for the 30 sets of experimental data:
\begin{equation}
\label{eq:distances}
d^2 = \sum_{k=1}^{30} \sum_{i=0}^{l-1} \left( P(N^\mathrm{det}_\mathrm{vox}=i) - \sum_{j=0}^{l-1} V_{ij} P(N_\mathrm{vox}=j) \right)^2,
\end{equation}
with the constraints $V_{ij} \geq 0$ and $\sum_i V_{ij} = 1$. We used a trust-region type algorithm from \texttt{scipy.optimize.minimize} to impose the constraints during the optimization \cite{SciPy, Byrd1999, Lalee1998}.
We have verified that the result of the optimization does not depend on the initial guess for the matrix $V$: identical optimized matrices $V$ are obtained when starting either from a random matrix, the identity matrix or a constant matrix.

Importantly, the optimisation procedure is performed by letting only the response of the detector, i.e. the coefficients $V_{ij}$, to vary. This is in contrast to the methods used in \cite{Hetzel2022, Keith2018} where both the statistics of the quantum state, i.e. the expected counting statistics $P(N_\mathrm{vox}=j)$, and the response of the detector ($V_{ij}$) are left to vary. Our choice stems from the fact that, in the case of our measurements, we find that the solution of the optimisation procedure is not unique if we leave both $P(N_\mathrm{vox}=j)$ and $V_{ij}$ to vary. For this reason, our prediction of the full counting statistics $P(N_\mathrm{vox}=j)$ in voxels (see Eq.~\eqref{eq:distribution}) is a crucial asset to perform a tomography of the single-atom-resolved detector. 

\begin{figure}[h!] 
\includegraphics[width=\columnwidth]{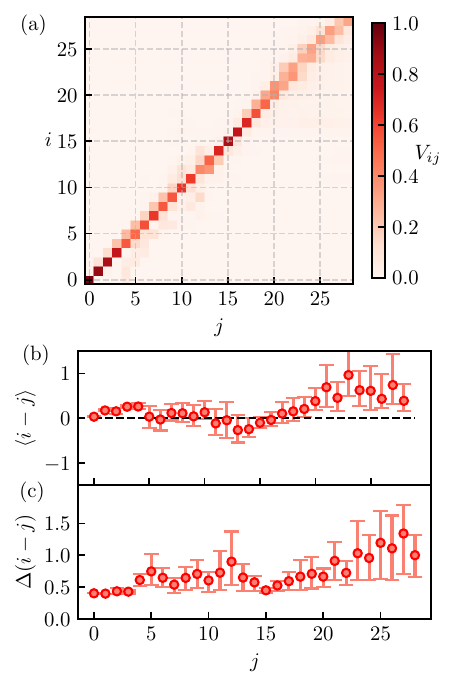}
\caption{\label{fig:Vij} (a) Reconstructed matrix $V=\{ V_{ij} \}$ of the He$^*$ detector from probing Mott insulators. (b)-(c) Averages $\langle i-j \rangle$ and standard deviations $\Delta (i-j)$ of the probabilities to detect $i-j$ atoms given that $j$ are expected. These values are obtained from fitting each column of the $V$ matrix with a normalized Gaussian function. Error bars correspond to 68\% confidence interval obtained from a bootstrap method and account for statistical fluctuations due to the finite sample size.}
\end{figure}

\subsection{Response of the MCP detector}

The matrix $V$ reconstructed using the data with Mott insulators is shown in Fig.~\ref{fig:Vij}(a). 
To avoid overfitting, we restrict the size of $V$ to $\sim 30 \times 30$ as we have 30 counting statistics.
We find that $V$ is approximately diagonal, indicating that the detector does not distort the measured full counting statistics. In addition, the background noise is negligible for our experimental parameters and the number of dark counts is $0.035(5)$.  
The broadening of the diagonal of $V$ at $j\gtrsim20$ is likely due to some underdetermination of the reconstruction occurring at large $j$ where fewer distributions have significant weights. 
In order to extract quantitative information on the response of our detector, we make use of the columns of the $V$ matrix, which quantify the probabilities to actually detect $i$ atoms while $j$ are expected. These probabilities are fitted by Gaussians to get a mean and standard deviation which correspond to the accuracy and precision at each atom number $j$, plotted in Fig.~\ref{fig:Vij}(b) and \ref{fig:Vij}(c). As expected from the matrix $V$, the fits confirm an accuracy at the single-atom level: $P(N_\mathrm{det}=i|N=j)$ is centered within the interval $]j-1, j+1[$. The precision at one standard deviation is $\sim\ 0.5$.

\section{Discussion of the tomography method}

\begin{figure}[h!] 
\includegraphics[width=\columnwidth]{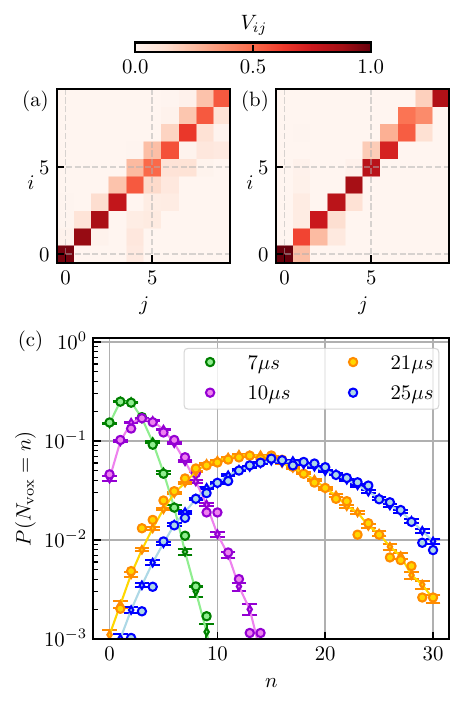}
\caption{\label{fig:V_comp} (a)-(b) Reconstructed matrix $V=\{ V_{ij} \}$ of the He$^*$ detector from probing Mott insulators, $V^\mathrm{Mott}$ (a) and BECs, $V^\mathrm{BEC}$ (b).
(c) Experimental counting statistics $P(N^\mathrm{det}_\mathrm{vox}=n)$ of the BEC data (dots) and the corresponding reconstructed counting statistics $\sum_j V^\mathrm{Mott}_{nj} P(N_\mathrm{vox}=j)$ using the matrix $V$ calibrated with Mott insulators (diamonds). Errorbars are 68\% confidence interval estimated using a bootstrap method.}
\end{figure}

To test our tomography method, we compare the results on Mott insulators, discussed above, with those obtained from probing Bose-Einstein condensates (BECs) which have different number statistics. We choose the voxel size such that the average number of detected particles $\langle N_\mathrm{vox} \rangle$ is similar to that used with Mott states. 
Because the density of a BEC in momentum space is much larger than that of a Mott insulator, the voxel size used in the BEC case is much smaller than that used in the case of a Mott insulator and not larger  than a correlation volume.
The bare statistics of BECs being Poissonian, the predictions of Eq.~\eqref{eq:algorithm params} still hold. We have recorded counting statistics at various Rabi durations and applied the optimization algorithm exactly as described previously.
The matrices $V$ reconstructed using the data with BECs and with Mott insulators are shown respectively in Fig.~\ref{fig:V_comp}(a) and Fig.~\ref{fig:V_comp}(b). We restrict the size of the matrices $V$ to $10 \times 10$ since we have 10 measured counting statistics with BECs. The  response $V$ obtained with BECs is close to be a diagonal matrix and agrees well with that obtained with Mott insulators. The consistency of the two independent reconstructions supports the validity of our tomography results.

As an alternative to performing a tomography with BECs, the experimental data using BECs can be compared to the reconstructed ones using the matrix $V = V_\mathrm{Mott}$ in order to test how tomography with a given state -- Mott insulators here -- can predict detection outcomes in general. By using the total atom number measured with BECs, we predict $P(N_{\rm vox}=n)$ from Eq.\eqref{eq:algorithm params}. The matrix $V=V^{\rm Mott}$ obtained through tomography with Mott insulators is then used to compute the expected counting statistics $P(N^\mathrm{det}_\mathrm{vox}=i)=\sum_{j=0}^{l-1} V^{\rm Mott}_{ij} P(N_\mathrm{vox}=j)$. Fig.~\ref{fig:V_comp}(c) shows a comparison between these reconstructed counting statistics for BECs and the actual measured ones, resulting in a remarkable agreement.

Two additional points regarding the efficiency of the detector are worth noting. First, the quantum efficiency $p$ of the detector is not accessible to our tomography method as both Poisson and thermal distributions are perfectly rescaled when affected by a binomial of efficiency $p$. The detection efficiency is implicitly included in the expected counting statistics when we use in Eq.\eqref{eq:algorithm params} the detected quantities $P(N^\mathrm{det}_\mathrm{tot}=\lambda)$ and $\alpha^\mathrm{det}$. The matrix $V$ determined above can be interpreted as gathering the probabilities to detect $i$ atoms given that we would have detected $j$ atoms with a detector whose response is described by a binomial distribution of value $p$. In other words, the matrix $V$ quantifies the difference between the response of our detector and that described by a binomial distribution. For completness, we determined $p=0.53(3)$ by comparing the average atom numbers detected by the MCP with those obtained by absorption imaging, see appendix \ref{app:C}. 
Second, we chose a regime of parameters such that the response of the detector is not affected by the saturation of the MCP. For larger atomic fluxes than those used in this work, the quantum efficiency would be reduced by a factor $1/(1+n/n_0)$, where $n_0 \sim 600$ is the threshold where saturation appears \cite{Fraser1991} and the matrix $V$ would deviate from a diagonal for numbers $n\geq n_{0}$.

Finally, we illustrate the effect of the calibrated detection on a quantum state whose properties are particularly sensitive to detection: a two-mode squeezed state. Our experiment produces two-mode squeezed states between modes of opposite momenta in the quantum depletion of Bose-Einstein condensates \cite{Tenart2021, Bureik2024}. 
For illustration, let us consider a perfect two-mode squeezed state $|\psi\rangle = \sum_n c_n |n\rangle_\mathbf{k} |n\rangle_{-\mathbf{k}}$. The variance of the number difference $\delta N_{k,-k}=N({\bm k})- N(-{\bm k})$ of the bare state (\textit{i.e.} before detection) is equal to zero, $\sigma_{\delta N}=\langle \delta N_{k,-k}^2 \rangle-\langle \delta N_{k,-k} \rangle^2=0$. If one accounts only for the detection efficiency $p=0.53(3)$ (see above) and describes the detection process with a binomial distribution, the measured variance would be significantly larger, $\sigma_{\delta N}^{\rm B}=\sqrt{2 p(1-p) \langle N({\bm k}) \rangle} = 3.9$ with $\langle N({\bm k}) \rangle=30$. Accounting for the reconstructed response of our detector, {\it i.e.} using the matrix $V$ on top of the binomial distribution, leads to $\sigma_{\delta N}^{\rm MCP}=4.3(1)$ for the same population $\langle N({\bm k}) \rangle=30$ of the modes ${\bm k}$ and $-{\bm k}$. This latter value is not significantly larger than that of $\sigma_{\delta N}^{\rm B}$, a result which highlights the fact that our detector is correctly described by a binomial distribution of value $p=0.53(3)$. The small difference ($\sim 10\%$) between $\sigma_{\delta N}^{\rm B}$ and $\sigma_{\delta N}^{\rm MCP}$ illustrates the sensitivity of the observable $\sigma_{\delta N}$ to the detection process and the interest in performing QDT.

\section{Conclusion}

We have presented a method to characterize a single-particle detector in presence of shot-to-shot number fluctuations. The tomography is performed in sub-volumes of the detector, typically larger than the correlation volume of the particles, while the counting statistics of the total particle number reaching the detector are used to quantify the role of shot-to-shot number fluctuations. The choices made within our method -- on quantum states and on voxel sizes -- allow us to predict the expected counting statistics, leading to a well-defined optimization algorithm from which the detector is characterized.
We have applied this method to characterize the POVMs of a 3D single-atom-resolved detector for metastable helium atoms. In the Fock state basis of \emph{detected} atom number, we find that the detector realises an almost perfect projection. This shows that the He$^*$ detector is overall well described by a binomial process with an efficiency $p=0.53(3)$. 
\\

{\bf Acknowledgements} We thank Thomas Chalopin and the members of the Quantum Gas group at Institut d’Optique for insightful discussions. We acknowledge financial support from the Région Ile-de-France in the framework of the DIM SIRTEQ, the “Fondation d’entreprise iXcore pour la Recherche”, the French National Research Agency (Grant number ANR-17-CE30-0020-01) and France 2030 programs of the French National Research Agency (Grant number ANR- 22-PETQ-0004).

\bibliography{biblio}

\appendix

\section{\label{app:A}Thermal statistics of the Mott insulator}

We defined the correlation volume $V_{c}=(2 l_{c})^3$ where $l_{c}$ is the half width at 1/e$^2$ of the two-body correlation function $g^{(2)}$ \cite{Herce2023}. In the case of Mott insulators, the correlation length is found to be isotropic (due to the presence of an isotropic harmonic trap) and is $l_{c} = 0.032(2) k_d$ where $k_d=2 \pi /d$ with $d = 775$ nm the lattice spacing. With this definition, atoms within the same volume of correlation are correlated.  

\begin{figure}[h!] 
\includegraphics{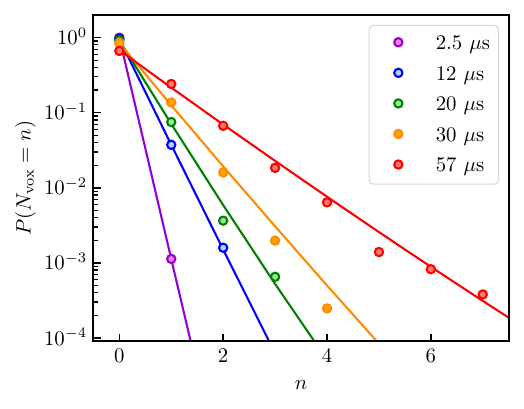} 
\caption{\label{fig:thermal distribution} Atom number distributions of the same data used in Fig.~\ref{fig:data}.b measured in voxels of volume $V=(0.025 k_d)^3 \sim 0.5 V_{c}$. Dots are experimental data and solid lines are the predictions from Eq.~\eqref{eq:thermal}.}
\end{figure}

After a long time of expansion from the trap, as implemented in our experiment, a bosonic Mott insulator exhibits a thermal statistics \cite{Carcy2019} which is reflected in the counting statistics measured in a volume smaller or equal to the volume of correlation. This counting statistics follow a thermal statistics (geometric distribution) with fluctuating parameter:
\begin{equation}
\label{eq:thermal}
\begin{split}
P(N_\mathrm{vox}=n) &= \sum_{\lambda \in \mathbb{N}} P(N_\mathrm{tot}=\lambda) P(N_\mathrm{vox}=n | N_\mathrm{tot}=\lambda)\\
    &= \sum_{\lambda \in \mathbb{N}} P(N_\mathrm{tot}=\lambda) \frac{1}{1+\alpha\lambda} \left( \frac{\alpha\lambda}{1+\alpha\lambda} \right)^n,
\end{split}
\end{equation}
where $\alpha = \langle N_\mathrm{vox} \rangle / \langle N_\mathrm{tot} \rangle$.
In Fig.~\ref{fig:thermal distribution}, we plot the counting statistics measured in a volume $V \sim 0.5 V_{c}$ for various pulse durations. The measured statistics agree well with the expected thermal distributions, see Eq.~\eqref{eq:thermal}. Recall that in the main text, the voxel we use is much larger than $V_c$, hence the expected Poisson statistics \cite{Herce2023}.

\section{\label{App:B} Homogeneity and averaging over similar voxels}

\begin{figure}[h!] 
\includegraphics{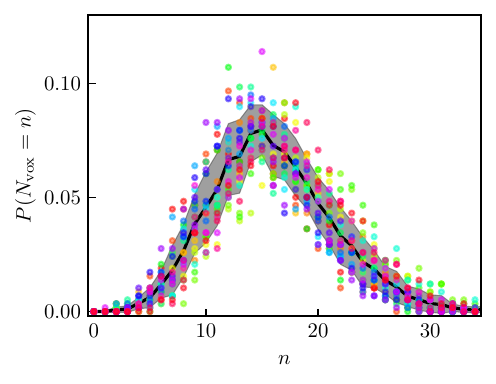}
\caption{\label{fig:homogeneity} Counting statistics in 24 voxels with identical sizes, each voxel is associated to a single color. The atom number is (mean $\pm$ standard deviation) $15.95\pm1.25$ atoms. The black line and gray area correspond to the mean and one standard deviation over these counting statistics, respectively.}
\end{figure}

In Fig.~\ref{fig:homogeneity}, we plot the counting statistics measured in the  individual voxels that we use to obtain the average counting statistics shown in the main text in Fig.~\ref{fig:distribution}. The counting statistics of individual voxels are identical up to the statistical noise. Additionally, the average counting statistics do not exhibit any broadening that would result from inhomogeneities between the different voxels. These results confirm the assumption that the response of our detector is homogeneous and they justify the average over several voxels used in the main text.

\section{\label{app:C} Finite detection efficiency, binomial response of the detector and Eq.~\eqref{eq:distribution}}

\begin{figure}[h!] 
\includegraphics[width=0.45\textwidth]{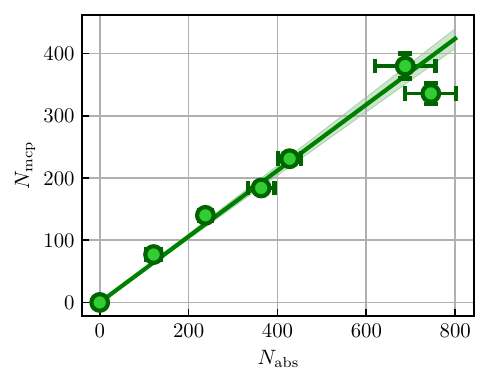} 
\caption{Measured atom number $N_\mathrm{mcp}$ as a function of the measured atom number $N_\mathrm{abs}$ with absorption imaging. The line is a linear fit whose slope $p=0.53(3)$ corresponds to the detection efficiency of the MCP detector.
\label{Fig:DetEff}}
\end{figure}

As discussed in the main text, we have performed a calibration of the detection efficiency of the MCP detector from a comparison with a calibrated absorption imaging. The results are plotted in Fig.~\ref{Fig:DetEff}. A linear fit yields a detection efficiency of $p=0.53(3)$.
\\

Equation~\eqref{eq:distribution} of the main text involves two quantities, $P(N_\mathrm{tot}=\lambda)$ and $\alpha$, for which we have not discussed the role of a finite detection efficiency $p$. In the experiment, we have access to the detected quantities only, $P(N^\mathrm{det}_\mathrm{tot}=\lambda)$ and $\alpha^\mathrm{det} = \langle N^\mathrm{det}_\mathrm{vox} \rangle / \langle N^\mathrm{det}_\mathrm{tot} \rangle$. If we account for the finite detection efficiency through using a binomial distribution with $p$ and if we assume that the detection noise $\propto \sqrt{Np(1-p)}$ is negligible compared to the shot-to-shot fluctuations, then we have
$$P(N_\mathrm{tot}=\lambda) \simeq P(N^\mathrm{det}_\mathrm{tot}=p\lambda) \mathrm{\ and\ } \alpha = \alpha^\mathrm{det}.$$
In our experiment, the measured shot-to-shot fluctuations have indeed a much larger distribution than that associated to the detection noise, so that the above assumption is verified.\\

The use of detected quantities in Eq.~\eqref{eq:distribution} implies that $P(N_\mathrm{vox}=n)$ is not the bare state statistics but the state statistics through a detection process described by a binomial response:
$$\begin{aligned}
P(N_\mathrm{vox}=n) &= \sum_{\lambda \in \mathbb{N}} P(N^\mathrm{det}_\mathrm{tot}=\lambda) P(N_\mathrm{vox}=n | N^\mathrm{det}_\mathrm{tot}=\lambda)\\
&= \sum_{\lambda \in \mathbb{N}} P(N^\mathrm{det}_\mathrm{tot}=\lambda) e^{-\alpha \lambda} \frac{(\alpha \lambda)^n}{n!}\\
&\simeq \sum_{\lambda \in \mathbb{N}} P(N_\mathrm{tot}=\lambda) \underbrace{e^{-\alpha p\lambda} \frac{(\alpha p\lambda)^n}{n!}}_\mathrm{Binomial\ of\ Poisson}
\end{aligned}$$

As discussed in the main text, the reconstructed matrix $V$ does not include the effect of the finite detection efficiency $p$ of the detector. To relate the measured quantities to the bare state statistics $P(N=k)$, we need to include this effect. These probabilities to detect $i$ atoms given that there are $k$ atoms can be obtained by multiplying $V$ by a matrix $B$ gathering the probabilities to expect the detection of $j$ atoms given that there are $k$ atoms:
\begin{equation}
\label{eq:full detection}
    P(N^\mathrm{det}=i) = \sum_k (V \times B)_{ik} \ P(N=k).
\end{equation}\\
where $B_{jk}=\binom{k}{j} p^j (1-p)^{k-j}$ describes a binomial distribution parametrized by $p$.

\end{document}